# Separation of Electron and Hole Dynamics in the Semimetal LaSb

F. Han[1,2], J. Xu[1,3], A. S. Botana[1], Z. L. Xiao*[1,3], Y. L. Wang[1,4], W. G. Yang[2], D. Y. Chung[1], M. G. Kanatzidis[1,5], M. R. Norman[1], G. W. Crabtree[1], and W. K. Kwok[1]

[1]Materials Science Division, Argonne National Laboratory, Argonne, Illinois 60439, USA
[2]Center for High Pressure Science and Technology Advanced Research, Shanghai 201203, China
[3]Department of Physics, Northern Illinois University, DeKalb, Illinois 60115, USA
[4]Department of Physics, University of Notre Dame, Notre Dame, Indiana 46556, USA
[5]Department of Chemistry, Northwestern University, Evanston, Illinois 60208, USA

We report investigations on the magnetotransport in LaSb, which exhibits extremely large magnetoresistance (XMR). Foremost, we demonstrate that the resistivity plateau can be explained without invoking topological protection. We then determine the Fermi surface from Shubnikov - de Haas (SdH) quantum oscillation measurements and find good agreement with the bulk Fermi pockets derived from first principle calculations. Using a semiclassical theory and the experimentally determined Fermi pocket anisotropies, we quantitatively describe the orbital magnetoresistance, including its angle dependence. We show that the origin of XMR in LaSb lies in its high mobility with diminishing Hall effect, where the high mobility leads to a strong magnetic field dependence of the longitudinal magnetoconductance. Unlike a one-band material, when a system has two or more bands (Fermi pockets) with electron and hole carriers, the added conductance arising from the Hall effect is reduced, hence revealing the latent XMR enabled by the longitudinal magnetoconductance. With diminishing Hall effect, the magnetoresistivity is simply the inverse of the longitudinal magnetoconductivity, enabling the differentiation of the electron and hole contributions to the XMR, which varies with the strength and orientation of the magnetic field. This work demonstrates a convenient way to separate the dynamics of the charge carriers and to uncover the origin of XMR in multi-band materials with anisotropic Fermi surfaces. Our approach can be readily applied to other XMR materials.

# I. INTRODUCTION

Magnetoresistance (MR), i.e., the change induced by a magnetic field in the electrical resistance [1], lies at the core of data storage in computer hard drives [2] and of other applications such as magnetic field sensors [3,4]. Since larger MRs can enhance the sensitivities of these devices, searching for new materials with large MRs has remained at the frontier of contemporary materials science research [5-23]. Besides the giant MR (GMR) [2] and colossal MR (CMR) [24] in magnetic thin films and compounds, extremely large MR (XMR) was observed decades ago in nonmagnetic materials such as bismuth [6] and graphite [7,8], though the underlying mechanism is still under debate [8,25-27]. The recent discovery of XMR in $PtSn_4$ [9], $PdCoO_2$ [10], $NbSb_2$ [11], and $WTe_2$ [12-14] and in particular, the revelation of XMR in exotic topological Dirac [15-17] and Weyl [18-23] semimetals have triggered extensive research to uncover its origin. More intriguingly, $PtSn_4$ [25] and $WTe_2$ [26] were also found to be topological semimetals, implying the possible relevance of topological protection for the observed XMR [27]. Other mechanisms such as a magnetic-field induced metal-insulator transition (MIT) [28-33], electron-hole ($e$-$h$) compensation [12,18,27,34], and forbidden backscattering at zero field [15] have also been considered as possible origins for XMR.

Recently, the rare-earth monopnictides LnX (Ln = La/Y/Nd/Ce and X = Sb/Bi) [27,34-49] were added to the family of XMR materials. These materials, with a rock-salt cubic crystal lattice, exhibit typical hallmark XMR behavior such as power-law magnetoresistance and magnetic-field induced turn-on behavior as a function of temperature. Due to their simple crystalline structure and possible topological nature [27,39,40], these materials represent good candidates for exploring the origin of XMR. The observed XMR in LaSb has been attributed to a magnetic-field-induced MIT [27,35], $e$-$h$ compensation [27], and high mobility of the Dirac-like bulk electronic bands [42].

Here, we aim to uncover the origin of the XMR in LaSb. Since previous transport [27,35] and ARPES [39] experiments on rare-earth monopnictides have raised the possible role of the surface states in the observed XMR, we first address the surface versus bulk issue in the magnetoresistance of our LaSb crystals. We demonstrate that the resistivity plateau at low temperatures, which was considered a signature of surface states [35], can be explained as a natural consequence of Kohler's rule scaling [50]. We show that the Fermi surfaces derived from our Shubnikov - de Haas (SdH) quantum oscillation measurements are in fact bulk Fermi pockets instead of surface ones. Furthermore, these materials have ellipsoidal electron Fermi pockets elongated along the Γ–X direction [27,35,36], which can result in an angle-dependent magnetoresistance [36,37,44-45] due to the anisotropic effective mass that governs the mobility of the charge carriers. Hence, a quantitative analysis of the angle dependence of the magnetoresistance may enable to reveal the role of each band / Fermi pocket on the XMR. We demonstrate that all bands /Fermi pockets play important roles in the observed XMR and their relative contributions vary with the strength and orientation of the magnetic field. This work also indicates that the popular isotropic two-band model with *e-h* compensation is insufficient to describe an anisotropic multi-band material.

We measured the resistivity of LaSb single crystals as a function of temperature as well as the strength and orientation of the magnetic field. The data can be quantitatively described with a semiclassical theory for an anisotropic system. We find that both electrons and holes have very high mobilities. More importantly, the magnetoresistivity is found to be nearly equal to the inverse of the longitudinal magnetoconductivity. This not only allows us to differentiate the contributions of electrons and holes but also to uncover the role of the Hall effect for the observed XMR: the latter controls the measurable portion of the XMR enabled by the longitudinal magnetoconductivity. The full potential of XMR is unmasked with a diminishing

Hall effect. That is, a high mobility with diminishing Hall effect is responsible for the occurrence of XMR in LaSb. In this multi-band material, every band plays an important role and the overall XMR reflects the contribution from all bands. We demonstrate that investigation on anisotropic magnetoresistance can provide a convenient way to separate the dynamics of the charge carriers and to uncover the origin of XMR in multi-band materials with anisotropic Fermi surfaces. This revealed mechanism can account for the XMR observed in other semimetals [9-14, 35-49].

## II. MATERIALS AND METHODS

**Crystal growth and characterization.** Single crystals of LaSb were synthesized in tin flux following the procedures in Ref.35. La powder (Alfa Aesar, 99.9%), Sb spheres (Alfa Aesar, 99.999%), and Sn pieces (Alfa Aesar, 99.999%) were loaded into an aluminum oxide crucible in a molar ratio of 1.5:1:20. The crucible with its top covered by a stainless steel sieve was then sealed in an evacuated silica ampule. The sealed ampoule was heated to 1050 °C over 10 hours, homogenized at 1050 °C for 12 hours and then cooled to 700 °C at the rate of 2 °C per hour. Once the furnace reached 700 °C, the tin flux was removed from the crystals using a centrifuge. Well-faceted crystals were collected on the stainless steel sieve. The crystal structure of the compound was verified by single crystal x-ray diffraction at room temperature using a STOE IPDS 2T diffractometer using Mo K$\alpha$ radiation ($\lambda$ = 0.71073 Å) and operating at 50 kV and 40 mA. The structure was solved by direct methods and refined by full-matrix least squares on $F^2$ using the SHELXTL program package [51].

**Resistivity Measurements.** We conducted DC resistivity measurements on two LaSb crystals (sample A and sample B) in a Quantum Design PPMS-9 using constant current mode ($I$ = 4 mA). The dimensions of the crystals are 223.33 μm ($w$) × 138.78 μm ($d$) × 790 μm ($l$) and 273.37 μm ($w$) × 218.93 μm ($d$) × 600 μm ($l$) for samples A and B, respectively. The electric contacts were made by attaching 50 μm diameter gold wires using silver epoxy, followed with baking at 120 °C for 20 minutes. In order to avoid sample degradation, the contacting operation was carried out in a glovebox with inert gas. Angular dependence of the resistance was obtained by placing the sample on a precision, stepper-controlled rotator with an angular resolution of 0.05°. The inset of Fig.1(a) shows the measurement geometry where the magnetic field $H(\theta)$ is rotated in the (100) plane and the current $I$ flows along the [100] direction, such that the magnetic field is always perpendicular to the applied current $I$. The resistivity versus temperature $\rho(T)$ curves at various magnetic fields were constructed by measuring $\rho(H)$ at various fixed temperatures. We define the magnetoresistance as $MR = [\rho - \rho_0)]/\rho_0$ where $\rho$ and $\rho_0$ are the resistivities at a fixed temperature with and without the presence of a magnetic field, respectively.

**First principles calculations.** The electronic structure calculations were carried out within density functional theory (DFT) using the all-electron, full potential code WIEN2K [52] based on the augmented plane wave plus local orbital (APW+lo) basis set [53]. The Perdew–Burke–Ernzerhof (PBE) version of the generalized gradient approximation (GGA) [54] was chosen as the exchange correlation potential. Spin-orbit coupling (SOC) was introduced in a second variational procedure [55]. A dense $k$-mesh of 34x34x34 was used for the Brillouin zone (BZ) sampling in order to check the fine details of the influence of spin-orbit coupling on the electronic structure. The product $R_{mt}K_{max}$ of the atomic sphere radius $R_{mt}$ and the plane wave cutoff parameter $K_{max}$ was chosen to be 7.0 for all the calculations. The $R_{mt}$ were 2.5 a.u. for both La and Sb.

### III. SURFACE VERSUS BULK TRANSPORT

Since LaSb was predicted to be a topological insulator [56], Tafti *et al* [35] attributed the XMR to surface states, backed by the observation of a resistivity plateau at low temperatures analogous to that in the topological insulator $SmB_6$ as well as the quasi-2D Fermi surfaces determined through quantum oscillation measurements. Assuming surface dominance in the conductance and using the lattice constant of 6.5 Å [27] as the thickness for the surface layer, we derived the sheet resistance for sample A at the plateau temperature regime to range from 1.2 mΩ at 1 T to 63 mΩ at 9 T. These values are well below the theoretically predicted sheet resistance (~30 Ω) for perfect graphene that has Dirac cones and is a ballistic conductor [57]. On the other hand, the surface in a topological material can have strong coupling with the bulk, enhancing the surface related conductance, e.g., by increasing the conductance of near-surface layers [39]. Below, we tackle the issue of surface versus bulk conductances in LaSb. We will show that bulk transport can in fact account for both the resistivity plateau and anisotropic Fermi surfaces, enabling us to confidently separate the dynamics of electrons and holes in the bulk, which is crucial to uncover the origin of the XMR.

### III.1. Understanding the resistivity plateau through Kohler's rule scaling

A signature of the XMR phenomenon is the so-called turn-on temperature behavior: when the applied magnetic field is above a certain value, the resistivity versus temperature $\rho(T)$ curve shows a minimum at a field dependent temperature $T_m$. For $T < T_m$, the resistivity increases dramatically with decreasing temperature while for $T > T_m$, it has a similar metallic temperature dependence as that in zero field. As presented in Fig.1(a) for sample A, the magnetoresistance of LaSb crystals displays the same temperature behavior: the $\rho(T)$ curve obtained at $\mu_0H = 1$ T or higher shows a dip at a field-dependent temperature. At very low temperatures ($T < 15$ K) the resistivity begins to saturate, forming a plateau in the $\rho(T)$ curve, as clearly shown in Fig.1(c) where the temperature is plotted in a logarithmic scale.

Since magnetic-field-induced MIT has been considered as a possible origin for the XMR in graphite [28], Tafti *et al* [35] attributed the turn-on temperature behavior in LaSb to a MIT. This MIT interpretation was adopted by other groups to account for the XMR in both rare-earth monopnictide [43] and other materials, including $NbAs_2$, $TaAs_2$, $TaSb_2$ and ZrSiS [31-33]. Below we use Kohler's rule scaling approach, which was successfully employed in $WTe_2$ by Wang *et al* [50], to reveal the origin for both the turn-on temperature behavior and the resistivity plateau. According to Wang *et al.*, $\rho(T)$ curves obtained for different magnetic fields follow Kohler's rule scaling $MR = \alpha(H/\rho_0)^m$, where $\alpha$ and m are sample dependent constants. We found that Kohler's rule scaling can account for the $\rho(T)$ relationship in LaSb, with $\alpha = 2.5 \times 10^{-10}$ $(\Omega cm/T)^{1.71}$ and $m = 1.71$, as shown in Fig.1(b). We can also re-write the Kohler's rule scaling as [50]:

$$\rho(T,H) = \rho_0 + \alpha H^m/\rho_0^{m-1} \qquad (1)$$

The second term is the magnetic-field-induced resistivity $\Delta\rho$. That is, the resistivity of a sample in a magnetic field consists of two components $\rho_0$ and $\Delta\rho$. Since $\Delta\rho \sim 1/\rho_0^{m-1}$, it has an opposite temperature dependence to that of the first term $\rho_0$. The competition of $\rho_0$ and $\Delta\rho$ with changing temperature results in a possible minimum at $T_m$ in the total resistivity $\rho(T,H)$. Fig.1(c) showcases how Eq.(1) can lead to the remarked turn-on behavior, where the resistivity at $\mu_0 H = 0$ T and $\mu_0 H = 9$ T as well as its difference $\Delta\rho = \rho(T,9T) - \rho_0$ are presented.

As demonstrated by Wang *et al* [50] for $WTe_2$, one can conveniently use Kohler's rule scaling to elucidate other turn-on behavior related features such as the magnetic field dependence of $T_m$ and the temperature dependence of the resistivity minima ($\rho_m$). Here, we show that Kohler's rule scaling can also describe the resistivity plateau, whose origin was also previously considered within a two-band model by Guo *et al* [34] and Sun *et al* [38]. As presented in Fig.1(c), the experimental data at $\mu_0 H = 9$ T, including the resistivity plateau, can be fitted well by Eq.(1) with derived values of $\alpha$ and *m* from Kohler's scaling and the experimentally obtained $\rho_0$. Since $\rho_0$ is the only temperature dependent variable in Eq.(1), the nearly perfect fits in Fig.1(c) indicate that the resistivity plateau originates from the temperature dependence of $\rho_0$. Following Eq.(1), at low temperatures $\Delta\rho \gg \rho_0$, thus $\rho(T,H) \approx \Delta\rho \sim 1/\rho_0^{m-1}$. Since $\rho_0$ is insensitive to temperature in the low temperature regime, a plateau is expected in the total resistivity. Hence, the resistivity plateau at high magnetic fields originates from the temperature-insensitive resistivity at zero-field. Since both the resistivity plateau and the turn-on temperature behavior can be derived using the same Kohler's rule scaling, they should represent behavior originating from the same region of the crystal, i.e. either the bulk or the surface states. This excludes the possibility that the turn-on behavior and the resistivity plateau [27,35] separately arise from the bulk and surface states of the crystal, respectively. Thus, we can safely conclude that the resistivity plateau in LaSb at low temperatures is a bulk property only.

**III.2. Revealing the bulk origin of the Shubnikov – de Haas Oscillations**

Although theory predicts that the rare-earth monopnictides can be topological insulators or semimetals [56], ARPES experiments from various groups have reached differing conclusions: multiple Dirac-like surface states near the Fermi level were observed in LaSb [39] and LaBi [39,40] and their odd number suggests these are topological materials, with Niu *et al* [39] concluding that the surface and near-surface bulk bands likely contribute strongly to the XMR in these two materials. But, Wu *et al* [41] found that the dispersion of the surface states resembles a Dirac cone with a linear dispersion for the upper band, separated by an energy gap from the lower band that follows roughly a parabolic dispersion instead. On the other hand, other ARPES results reveal that both LaSb [42] and YSb [46] are topologically trivial, as they did not observe surface states, with a bulk band structure consistent with band theory. For a topological material whose surfaces could be very conductive, its electrical conductance can come from both the bulk and the surface states.

We conducted both angle dependent SdH oscillations measurements and first principles calculations on LaSb. In previous SdH oscillation experiments on LaSb [27,35] and LaBi [36], the current flowed in the rotation plane of the magnetic field. Such a configuration may lead to ambiguity in determining the oscillation feature when the field direction is near or parallel to that of the current, where the Lorentz force is weak or diminishes. To mitigate this effect, in our experiments the current flow is perpendicular to the field rotation plane as shown by the schematic in the inset of Fig.1(a) and hence the Lorentz force and the orbital magnetic field remains unchanged under varying magnetic field orientation [48].

Figure 2(b) shows a typical $\rho(H)$ curve at a low temperature ($T$ = 2.5 K) and at a specific magnetic field orientation ($\theta$ =121°). SdH oscillations can be seen at high fields. The inset of Fig.2(b) shows the oscillations after subtracting a smooth background from the $\rho(H)$ curve. The amplitude of the oscillations does not decrease monotonically with decreasing field as demonstrated in Fig. 2(b). Instead the observed beating behavior indicates that the oscillations contain more than one frequency. If the oscillations originated from the surface states, we would expect to see up to two frequencies in case the side surfaces of the crystal do not have exactly the same states as those of the top/bottom surfaces. However, FFT analysis shown in Fig.2(c) reveals more than two frequencies. Furthermore, FFT results presented in Fig.2(d) over a wide range of field orientations show additional frequencies than those that can be described by the angle dependence of the SdH oscillation frequency of 2D Fermi surfaces, $F \sim 1/\cos(\theta\text{-}n\pi/2)$ [27,35,36]. Thus, we need to include the contributions from the bulk.

The bulk electronic band structure and Fermi surface of LaSb were investigated more than three decades ago [58-60] and also reported in recent publications [27,35,42]. As shown in the projection on the $k_y$-$k_z$ plane in Fig.2(a) and in the 3D plot in Fig.3(b), the bulk Fermi surface consists of electron pockets centered at X and elongated along the $\Gamma$–X direction in addition to two hole pockets centered at $\Gamma$. That is, we could observe up to five fundamental frequencies from the bulk Fermi surface with our current-magnetic field configuration given in the inset of Fig.1(a). In Fig.2(c) we indeed can identify four fundamental frequencies and their higher harmonics. Applying the same analysis procedure to the frequencies of the SdH oscillations obtained at other angles, we derive the angle dependences of the four fundamental frequencies. The results are presented as dark blue, green, red and purple solid circles in Fig.2(e) and labeled as $F\alpha_1$, $F\alpha_2$, $F\beta$ and $F\gamma$ (where $\alpha$ denotes the electron surfaces and $\beta$, $\gamma$ the hole surfaces). Their higher harmonics are presented with the same symbols but with lighter colors.

Although we obtained exactly the same frequency of 212 Tesla for $H /\!/$ [001] and $H /\!/$ [010] as previously reported [27,35], the anisotropy of ~ 4 for $F\alpha_1$, $F\alpha_2$ and the observation of $F\beta$ and $F\gamma$ exclude 2D surface states to be (solely) responsible for the observed SdH oscillations. On the other hand, the anisotropy for $F\alpha_1$, $F\alpha_2$ is nearly the same as that of the bulk electron pockets revealed by ARPES [42] and dHvA oscillation [58,59] experiments.

In order to better understand the data in Fig.2(e), we calculated the angle dependence of the SdH oscillation frequencies from band theory for the bulk Fermi pockets. As presented in Fig.3(c) and Fig.3(d), the $\alpha_1$ and $\alpha_2$ electron pockets and the $\beta$ and $\gamma$ hole pockets produce SdH oscillations with angle dependences very close to those in Fig.2(e): we obtain an anisotropy of 4.25 and a minimum frequency of 225 T for the electron pockets. The calculated angle dependences for all the $\alpha_1$, $\alpha_2$, $\beta$ and $\gamma$ pockets agree well with experimental data, as shown in Fig.3(e), where the calculated frequencies are multiplied with a scaling factor close to 1, indicating a slight deficiency of the DFT-derived Fermi surface.

We note that the frequencies expected for the electron pocket $\alpha_3$ in Fig.3(c) could not be identified from the experimental data in Fig.2(e), similar to that found in the dHvA data [58,59]. One interpretation for this absence is that the frequencies for the electron pocket $\alpha_3$ in LaSb are about twice that of the hole pocket β and hence hidden by the second harmonic of the latter. On the other hand, recent work on YSb revealed an alternative explanation [48]: the current flows along the long axis of the elliptical $\alpha_3$ Fermi pocket, and hence the mobility of the associated electrons is low [see discussions below: the mobility ($\mu_\parallel$) of the electrons from the $\alpha_3$ Fermi pocket is a factor of ~16 ($=\lambda_\mu^2$) smaller than that ($\mu_\perp$) of the $\alpha_1$ and $\alpha_2$ Fermi pockets]. Since the oscillation amplitude depends exponentially on the mobility, $\Delta\rho \sim e^{-1/\mu H}$, the SdH quantum oscillations from the $\alpha_3$ Fermi pocket could be below the measurement sensitivity level associated with our maximum magnetic field of 9 Tesla.

Based on the above discussions we can confidently conclude that the SdH oscillations observed in our LaSb crystals are solely from the 3D bulk Fermi surfaces, with $F_{\alpha 1}$, $F_{\alpha 2}$, $F_\beta$ and $F_\gamma$ from the electron pockets α₁ and α₂ and hole pockets β and γ, respectively. In Fig.2(e) all the detectable frequencies from the FFT analysis can be assigned to these four fundamental frequencies and their higher harmonics. We see that some frequencies can also originate from more than one Fermi pocket.

The experimental data presented in Fig.2(e) yields a complete picture of the anisotropy of the bulk Fermi surface in LaSb: the electron pockets are highly anisotropic while the hole pockets are nearly isotropic. Quantitatively, the angle dependence of $F_{\alpha 1}$, $F_{\alpha 2}$ can be fit with

$$F_\alpha = F_0/\sqrt{\cos^2[\theta - (n-1)\pi/2] + \lambda_\mu^{-2}\sin^2[\theta - (n-1)\pi/2]}$$

where $F_0$ = 211.5 Tesla, $\lambda_\mu$ = 4.1, and $n$ = 1, 2 for the α₁, α₂ pockets, respectively.

Since the oscillation frequency $F$ is proportional to the extremal orbit area $A = \pi(k_F^S)^2/\sqrt{\cos^2\theta + \lambda_\mu^{-2}\sin^2\theta}$ with $\lambda_\mu = k_F^L/k_F^S$ and $k_F^L$ and $k_F^S$ being the semimajor and semiminor axes of the elliptic Fermi pocket. Using the Onsager relation $F = (\phi_0/2\pi^2)A$ with $\phi_0$ being the flux quantum [27], we obtain the short Fermi vector $k_F^S = (2\pi F_0/\phi_0)^{1/2} = 8.02 \times 10^6$ cm$^{-1}$ and the long Fermi vector $k_F^L = \lambda_\mu k_F^S = 3.288 \times 10^7$ cm$^{-1}$ for the electron ellipsoid, corresponding to a density of $7.134 \times 10^{19}$ cm$^{-3}$ for each pocket and a total electron density of $2.14 \times 10^{20}$ cm$^{-3}$. This value is significantly larger than the previously reported one ($1.6 \times 10^{20}$ cm$^{-3}$) [27], although the values of their $F_0$ and the short $k_F$ are nearly the same as ours. Note that we used $n = k_F^L(k_F^S)^2/3\pi^2$ to calculate the density rather than the typically used $n = k_F^3/3\pi^2$ for a spherical Fermi surface [15].

The frequencies for the two hole pockets show a slight angle dependence with a four-fold symmetry. Mathematically, we can fit the data for the $\beta$ and $\gamma$ pockets respectively with

$$F_\beta = 430/\sqrt{\cos^2 2\theta + 1.04^{-2}\sin^2 2\theta} \quad \text{and}$$

$$F_\gamma = 890/\sqrt{\cos^2(2\theta - \pi/2) + 1.23^{-2}\sin^2(2\theta - \pi/2)}$$

The 'anisotropy' of 1.04 and 1.23 for the $\beta$ and $\gamma$ pockets is much smaller than that (4.1) of the $\alpha$ pockets. To calculate the hole density we treat both $\beta$ and $\gamma$ pockets as spheres with the average frequencies of 438 T and 995 T, corresponding to hole densities of $5.186 \times 10^{19}$ cm$^{-3}$ and $1.774 \times 10^{20}$ cm$^{-3}$, respectively. Thus, the total hole density is $2.29 \times 10^{20}$ cm$^{-3}$, which is ~6% higher than the electron density, indicating that LaSb is indeed a nearly compensated semimetal, consistent with the ARPES finding on the electron-hole ratio [42].

## IV. SEPARATION OF THE BULK ELECTRON AND HOLE DYNAMICS

Figure 4(a) presents the magnetic field dependence of the sample resistivity $\rho(H)$ at $T = 3$ K and $H$ // [001]. As shown in the inset, this sample has a large MR of $4.45 \times 10^4$ % at $\mu_0 H = 9$ T. At high magnetic fields, $\rho(H)$ follows a power-law dependence with an exponent of 2, consistent with that reported in other XMR materials [12, 50]. Such a quadratic relationship that implies a non-saturating magnetoresistance can be derived from the isotropic two-band model with $e$-$h$ compensation [12], which has become the most prevalent explanation for the origin of the XMR [12,27,34,36,37]. Indeed, ARPES experiments [42] and our SdH oscillation measurements (section III.2) reveal a nearly perfect $e$-$h$ compensation in LaSb. We can also use the two-band model to fit $\rho(H)$ of our LaSb crystal, with the derived physical parameters ($n_e$ = 9.03 × 10$^{19}$ cm$^{-3}$; $n_h$ = 8.77 × 10$^{19}$ cm$^{-3}$; $\mu_e$ = 0.673 m$^2$V$^{-1}$s$^{-1}$; $\mu_h$ = 0.639 m$^2$V$^{-1}$s$^{-1}$) very close to those reported in Ref.27. Although the $n_e/n_h$ ratio does indicate a nearly compensated nature, the absolute values of the $n_e$ and $n_h$ are less than half of those determined by the SdH experiments in section III.2. Furthermore, the isotropic two-band mode cannot account for the four-fold angle dependence of the resistivity $\rho(\theta)$, as delineated in Fig.4(b). Although the surface states of a topological material such as SmB$_6$ could induce a similar four-fold angular magnetoresistance [61,62], the anisotropy in Fig.4(b) shows that nearly perfect four-fold symmetry should not arise from the crystal surfaces, since the crystal's width (223.33 μm) is much larger than its thickness (138.78 μm). ARPES experiments [42] have revealed that LaSb is topologically trivial. Also, analysis on the resistivity plateau and quantum oscillations in section III clearly revealed their bulk origin. The magnetoresistance of a material with a bulk anisotropic Fermi surface can also vary with magnetic field orientation due to the anisotropic mobility [63,64]. Strong anisotropy in the magnetoresistance was observed in XMR materials such as bismuth [63,64] and graphite [65] as well as in WTe$_2$ [66]. As presented in Fig.2(a), which shows the projection of the calculated Fermi pockets in the magnetic field rotation plane), LaSb has two pairs of elongated electron Fermi pockets $\alpha_1$ and $\alpha_2$ in the $k_y$-$k_z$ plane, and hence an anisotropic magnetoresistance is expected when the magnetic field is rotated in this plane. In order to correctly describe the magnetoresistance anisotropy, we need to consider the effects of the elongated Fermi pockets.

For a non-topological material with a bulk ellipsoidal electron Fermi surface, the longitudinal magnetoresistivity with current flowing along the third axis is given as follows [64]:

$$\rho_{33} = \frac{1}{\sigma_{33} + \delta\sigma_{33}} \qquad (2)$$

where $\sigma_{33}$ and $\delta\sigma_{33}$ are the respective longitudinal magnetoconductivity and the additional magnetoconductivity induced by the Hall effect in the current flowing direction, given as:

$$\delta\sigma_{33} = \frac{\sigma_{12}\sigma_{23}\sigma_{31} + \sigma_{13}\sigma_{21}\sigma_{32} - \sigma_{11}\sigma_{23}\sigma_{32} - \sigma_{22}\sigma_{13}\sigma_{31}}{\sigma_{11}\sigma_{22} - \sigma_{12}\sigma_{21}} \qquad (3)$$

$$\sigma_{ij} = \frac{\sigma_{ij}^{e0}}{1 + H^2\mu_3[\mu_2 cos^2\theta + \mu_1 sin^2\theta]} \qquad (4)$$

where $\sigma_{11}^{e0} = ne(\mu_1 + H^2 cos^2\theta\mu_1\mu_2\mu_3)$; $\sigma_{22}^{e0} = ne(\mu_2 + H^2 sin^2\theta\mu_1\mu_2\mu_3)$; $\sigma_{33}^{e0} = ne\mu_3$; $\sigma_{12}^{e0} = \sigma_{21}^{e0} = neH^2 cos\theta sin\theta\mu_1\mu_2\mu_3$; $\sigma_{13}^{e0} = -\sigma_{31}^{e0} = -neH sin\theta\mu_1\mu_3$; and $\sigma_{23}^{e0} = -\sigma_{32}^{e0} = neH cos\theta\mu_2\mu_3$. Here $n$ is the electron density, $\mu_1$, $\mu_2$ and $\mu_3$ are the mobilities along the three axes of the ellipsoid. The magnetic field rotates in the 1-2 plane and $\theta$ is the angle of the magnetic field tilted away from the first axis. Eqs.(2)-(4) are applicable to an ellipsoidal hole pocket by changing the sign of both the charge $e$ and the mobility. They can also be implemented for the case of a spherical Fermi pocket by assuming $\mu_1 = \mu_2 = \mu_3$. Eq.(4) indicates that the magnetoconductivity of each Fermi pocket is hence determined by four parameters ($n$, $\mu_1$, $\mu_2$ and $\mu_3$).

LaSb has one electron band $\alpha$ with three orthogonal Fermi pockets ($\alpha_1$, $\alpha_2$ and $\alpha_3$) and two hole bands $\beta$ and $\gamma$ (see Fig.2(a) and Fig.3(b)). In order to account for the measured magnetoresistivity, we need to include contributions from all five Fermi pockets, i.e. to replace the $\sigma_{ij}$ in Eq.(4) with $\sigma_{ij}^T = \sum_k \sigma_{ij}^k$, where $k = \alpha_1$, $\alpha_2$, $\alpha_3$, $\beta$ and $\gamma$. Once the ratio of the ellipsoid's semimajor and semiminor axes $k_F^L$ and $k_F^S$ is known, the relationship of the mobility along the long axis $\mu_{//}$ and the short axis $\mu_\perp$ can be described as $\mu_\perp/\mu_{//} = m_{//}/m_\perp = (k_F^L/k_F^S)^2 = \lambda_\mu^2$, where $m_{//}$ and $m_\perp$ are the effective masses along the long and short axes [65]. That is, only one of the three mobilities is an independent fitting parameter. For example, we can rewrite the longitudinal magnetoconductivity for the $\alpha_1$, $\alpha_2$ and $\alpha_3$ Fermi pockets as $\sigma_{33}^{e1} = \sigma_{33}^{10}/[1 + \mu_\perp^2 H^2(\cos^2\theta + \sin^2\theta/\lambda_\mu^2)]$, $\sigma_{33}^{e2} = \sigma_{33}^{20}/[1 + \mu_\perp^2 H^2(\cos^2\theta/\lambda_\mu^2 + \sin^2\theta)]$, and $\sigma_{33}^{e3} = \sigma_{33}^{30}/[1 + \mu_\perp^2 H^2/\lambda_\mu^2]$, with $\sigma_{33}^{10} = n_{e1}e\mu_\perp$, $\sigma_{33}^{20} = n_{e2}e\mu_\perp$ and $\sigma_{33}^{30} = n_{e3}e\mu_{//}$ to be the zero-field conductivity for $\alpha_1$, $\alpha_2$ and $\alpha_3$ electron pocket, respectively. $\lambda_\mu$ can be determined through Shubnikov – de Haas (SdH) quantum oscillation measurements, with $\lambda_\mu = 4.1$ for sample A. Due to the crystalline symmetry, $\alpha_1$, $\alpha_2$ and $\alpha_3$ are exactly the same Fermi pocket but oriented differently. Thus, they have the same electron density, i.e., $n_{e1} = n_{e2} = n_{e3}$, leading to $\sigma_{33}^{10} = \sigma_{33}^{20} = \sigma_{33}^{e0}$ and $\sigma_{33}^{30} = \sigma_{33}^{e0}/\lambda_\mu^2$, where $\sigma_{33}^{e0} = n_{e1}e\mu_\perp$. That is, we have only two independent fitting parameters ($\sigma_{33}^{e0}$ and $\mu_\perp$) for the three electron Fermi pockets. For simplification we combine the two spherical hole bands into one with an isotropic mobility of $\mu_H$ and a zero-field conductivity of $\sigma_{33}^{h0}$. With these two parameters we can obtain $\sigma_{ij}$ for the combined hole bands, e.g., $\sigma_{33} = \sigma_{33}^{h0}/[1 + \mu_H^2 H^2]$.

As presented in Fig.4, Eq.(2) can quantitatively describe the measured $\rho(H)$ and $\rho(\theta)$ with $\sigma_{33}^{e0} = 6.05 \times 10^5$ Scm$^{-1}$, $\sigma_{33}^{h0} = 3.0 \times 10^5$ Scm$^{-1}$, $\mu_\perp = 9.27$ m$^2$V$^{-1}$s$^{-1}$ ($\mu_{//} = \mu_\perp/\lambda_\mu^2 = 0.552$ m$^2$V$^{-1}$s$^{-1}$), and $\mu_H = 1.5$ m$^2$V$^{-1}$s$^{-1}$. Interestingly, the above hole mobility is not far away from that ($\mu_h = 0.964$ m$^2$V$^{-1}$s$^{-1}$) derived using the isotropic two-band model that gives an electron mobility ($\mu_e = 1.118$ m$^2$V$^{-1}$s$^{-1}$) differing significantly from the values of $\mu_\perp$ and $\mu_{//}$.

Eq.(2) not only depicts a system with anisotropic Fermi pockets, it also provides a convenient way to reveal the role played by the Hall effect in the occurrence of XMR, as demonstrated in Fig.5. Using the parameters obtained from fitting $\rho(H)$ in Fig.4(a) with Eq.(2), we can calculate $\sigma_{33}$ and $\delta\sigma_{33}$ for each Fermi pocket and combinations thereof. If only the $\alpha_1$ Fermi pocket were present, the Hall effect induced additional magnetoconductivity $\delta\sigma_{33}$ would perfectly compensate $\sigma_{33}$ at any given magnetic field. This leads to an unchanging $\rho_{33}$ with varying magnetic field and hence the absence of a MR, since the magnetoresistance reflects the deflection of charge carriers by the magnetic field. The MR vanishes when the Hall field completely compensates for the deflection produced by the magnetic field. The situation remains the same even with the addition of the $\alpha_2$ Fermi pocket that has the same mobility ($\mu_\perp$) as that of $\alpha_1$ in the current flowing direction. However, the addition of the $\alpha_3$ Fermi pocket with a different mobility ($\mu_{//}$) reduces $\delta\sigma_{33}$, resulting in a magnetic field dependent $\rho_{33}$ with a $MR \approx 330\%$ at $\mu_0 H = 9$ T. This is because the Hall field cannot prevent the different mobility electrons from being deflected by the magnetic field, resulting in a finite magnetoresistance [1]. By further adding hole pockets, $\delta\sigma_{33}$ becomes negligible and the magnetic field dependence of $\rho_{33}$ follows $1/\sigma_{33}$. That is, in our LaSb crystal for $H // [001]$, the contribution of the Hall effect to the total magnetoconductivity nearly vanishes, leading to $\rho_{33} \approx 1/\sigma_{33}$. Fig.6 shows that this conclusion is valid for all magnetic field orientations. These discussions reveal that the origin of XMR comes from a high mobility with diminishing Hall effect. High mobility accounts for the strong magnetic field dependence of the longitudinal magnetoconductivity $\sigma_{33}$. A diminishing Hall effect that gives rise to $\rho_{33} \approx 1/\sigma_{33}$ enables $\rho_{33}$ to benefit from the drastic reduction of $\sigma_{33}$ with increasing magnetic field, leading to XMR. In other words, $\sigma_{33}$ determines the upper limit of the MR (denoted as $MR\sigma$ in Fig.5), which can be completely unmasked when the Hall field vanishes. Fig.5(c) even indicates that a large MR can occur in materials with only one type of charge carrier with different high mobilities, although the values could saturate at high magnetic fields.

The diminishing Hall effect, which leads to $\rho_{33} \approx 1/\sigma_{33}$, enables us to uncover the roles played by each type of charge carrier and/or even each band in the observed XMR. In Fig.7(a), we present the separation of electron and hole $\rho(H)$ behavior for $H//[001]$. It indicates that the electrons are more conductive at low fields but becomes nearly the same as that of the holes at high fields. The $\rho(\theta)$ presented in Fig.7(b) for $\mu_0 H = 9$ T shows that the electrons have a larger resistivity at nearly all field orientations except along the two principal axes of the crystal. That is, the contribution of the electrons and holes to the measured magnetoresistivity varies with the orientation of the magnetic field. We can also determine each electron pocket's contribution to the magnetoresistivity for any strength and orientation of the magnetic field. In Figs.7(c) and (d), we plot $\rho(H)$ for each of the electron pockets and their sum for the [001] direction and $\rho(\theta)$ at $\mu_0 H = 9$ T, respectively. We can see that the $\alpha_1$ and $\alpha_2$ pockets are the major contributors to the total electron conductivity, with nearly all coming from the $\alpha_2$ pocket in the [001] direction for $H > 1.0$ T.

Based on the above analysis, we can obtain the field and angle dependences of the magnetoresistance MR for each type of charge carrier and each electron pocket. Examples of $MR(H)$ for $H//[001]$ and $MR(\theta)$ for $\mu_0 H = 9$ T are presented in Fig.8, which indicates that: (1) both holes and electrons as well as each electron pocket exhibit XMR, (2) electrons have larger MR than holes for all field orientations, (3) the total (measured) MR can have a magnetic field dependence different from a power-law relationship with an exponent of two (see Fig.8(a)) even if the electrons in each Fermi pocket and the holes have quadratic MR.

## V. CONCLUSION

In summary, we succeeded in the separation of electron and hole dynamics in LaSb by investigating the anisotropic magnetoresistance and demonstrated that the XMR in LaSb crystals originates solely from the bulk. We used Kohler's rule scaling to understand the observed resistivity plateau without having to invoke topological protection. We conducted Shubinkov – de Haas oscillation experiments and found that the results agree well with our analysis of the bulk Fermi surfaces, excluding a possible surface origin. We further showed that both the magnetic field and angle dependences of the sample resistivity can be quantitatively described with a semiclassical theory that accounts for the anisotropic mobility of the ellipsoidal electron Fermi pockets. The analysis indicates that both the electrons and holes have high mobility and the multi-band nature results in a diminishing Hall effect. The high mobility together with diminishing Hall effect lead to the observed XMR in LaSb: the high mobility produces a strong field dependent longitudinal magnetoconductivity. With diminishing Hall effect, the measured magnetoresistivity becomes the inverse of the longitudinal magnetoconductivity, leading to the emergence of XMR behavior. Both the electrons and holes are found to play important roles in the observed XMR and their relative contributions vary with the strength and orientation of the magnetic field. We demonstrated that investigations of the anisotropic magnetoresistance provide a convenient way to separate the dynamics of charge carriers and to uncover the origin of the XMR in multi-band materials with anisotropic Fermi surfaces. The high mobility with diminishing Hall effect can also explain XMR behavior in other materials.

**ACKNOWLEDGEMENTS**

Crystal growth, vector-magnetotransport measurements and band structure calculations were supported by the U.S. Department of Energy, Office of Science, Basic Energy Sciences, Materials Sciences and Engineering Division. Rotational magnetoresistance experiments were carried out at Northern Illinois University under Grant No. DMR-1407175. Electrical contacting was conducted at Argonne's Center for Nanoscale Materials (CNM), which was supported by DOE BES under Contract No. DE-AC02-06CH11357. Structural characterization were supported by the National Nature Science Foundation of China under Contract No. U1530402.

*Corresponding author, xiao@anl.gov or zxiao@niu.edu

**Figure captions**

**FIG.1.** (a) Temperature dependence of the resistivity $\rho(T)$ of sample A at various magnetic fields. Inset: schematic showing the definition of the angle $\theta$ for the magnetic field orientation. The magnetic field rotates in the (100) plane while the current flows in the [010] direction, i.e. they are always perpendicular to each other. (b) Kohler's rule scaling of $\rho(T)$ presented in (a). The symbols are the experimental data and the solid line is a fit to $MR \sim (H/\rho_0)^{1.71}$. (c) Temperature dependence of the resistivity at $\mu_0H = 0$ T and $\mu_0H = 9$ T and their differences. The solid lines are fits to Eq.(1) with $\alpha = 2.5 \times 10^{-10}$ [$\Omega$cm/T]$^{1.71}$ and $m = 1.71$.

**FIG.2.** (a) Projection of the calculated Fermi pockets in the magnetic field rotation plane – the (100) plane. (b) A typical resistance versus magnetic field curve obtained for sample A at low temperatures, which shows Shubnikov - de Haas (SdH) oscillations at high magnetic fields. The inset is the oscillatory component after subtracting a smooth background. (c) Fast Fourier transform (FFT) analysis of the SdH oscillation in the inset of (b). (d) FFT amplitude versus frequency curves for field orientations at angles from $\theta = 91°$ to $271°$ in step sizes of $10°$. (e) Angle dependence of the SdH oscillation frequencies. Symbols are experimental data and lines are fits to the equations described in the text of the angle dependences for F$\alpha$, F$\beta$ and F$\gamma$. The darker symbols and lines represent the fundamental frequencies assignable to the Fermi pockets $\alpha_1$, $\alpha_2$, $\beta$ and $\gamma$. The lighter symbols and the dashed lines are their corresponding higher harmonics (up to the fourth one for the $\alpha_1$, $\alpha_2$ and $\beta$ with a few second harmonics for $\gamma$). The bi-colored symbols represent frequencies that could result from two different Fermi pockets (see text).

**Fig.3.** (a) Electronic band structure for LaSb obtained from first principles calculations. (b) Three-dimensional (3D) plot of the calculated Fermi surface. (c) Angle dependence of the calculated SdH oscillation frequencies in the (100) plane for all five Fermi pockets. (d) and (e) Comparison of the experimental (symbols) and the calculated values. In (e) the calculated curves are scaled by the factor listed and presented as dashed lines to show that their angle dependences are the same as those of the experimental data.

**FIG.4.** Magnetic field (a) and angle (b) dependences of the resistivity of sample A at $T = 3$ K. Symbols are the experimental data and solid lines are fits to Eq.(2). Inset of (a) shows the *MR*, with the dashed line presenting a quadratic field dependence.

**FIG.5.** Demonstration of the role of the Hall effect on the occurrence of XMR. $\sigma_{33}$, $\delta\sigma_{33}$, $\rho_{33}$ are calculated using Eqs.(2)-(4) with parameters derived from fitting the data in Fig.4(a) for Fermi pocket $\alpha_1$ (a), $\alpha_1+\alpha_2$ (b), $\Sigma\alpha = \alpha_1+\alpha_2+\alpha_3$ (c), and all Fermi pockets $\Sigma\alpha+\beta+\gamma$ (d). The inset in each panel is the projection of the corresponding Fermi pockets in the magnetic field rotation plane. $MR\sigma$ and $MR\rho$, which denote the MRs calculated from the magnetic field dependence of $1/\sigma_{33}$ and $\rho_{33}$, are the potential MR and measurable MR, respectively. In (a) and (b) the $MR\rho$ curves cannot be seen in logarithmic coordinates because $\rho_{33}$ is independent of magnetic field, resulting in $MR\rho = 0$.

**Fig.6.** Comparison of the calculated angle dependence of the total $\rho_{33}$ (red lines) and $1/\sigma_{33}$ (blue lines) at $T = 3$ K and various magnetic fields.

**FIG.7.** Separation of the magnetoresistivity $\rho(H, \theta)$ of sample A at $T = 3$ K: (a) and (b) are the field dependence for $H \;//\; [001]$ and angle dependence at $\mu_0H = 9$ T for the electrons and holes as well as their sum, respectively. (c) and (d) show the field dependence for $H \;//$

[001] and angle dependence at $\mu_0 H = 9$ T for the three electron pockets and the total electron value, respectively.

**FIG.8.** Separation of the magnetoresistance MR of sample A at $T = 3$ K for the data in Fig.4: (a) and (b) are the field dependence for $H \mathbin{/\mkern-2mu/} [001]$ and angle dependence at $\mu_0 H = 9$ T for the electrons and holes as well as their sum, respectively. (c) and (d) show the field dependence for $H \mathbin{/\mkern-2mu/} [001]$ and angle dependence at $\mu_0 H = 9$ T for the three electron pockets and the total electron value, respectively.

**FIGURE 1**

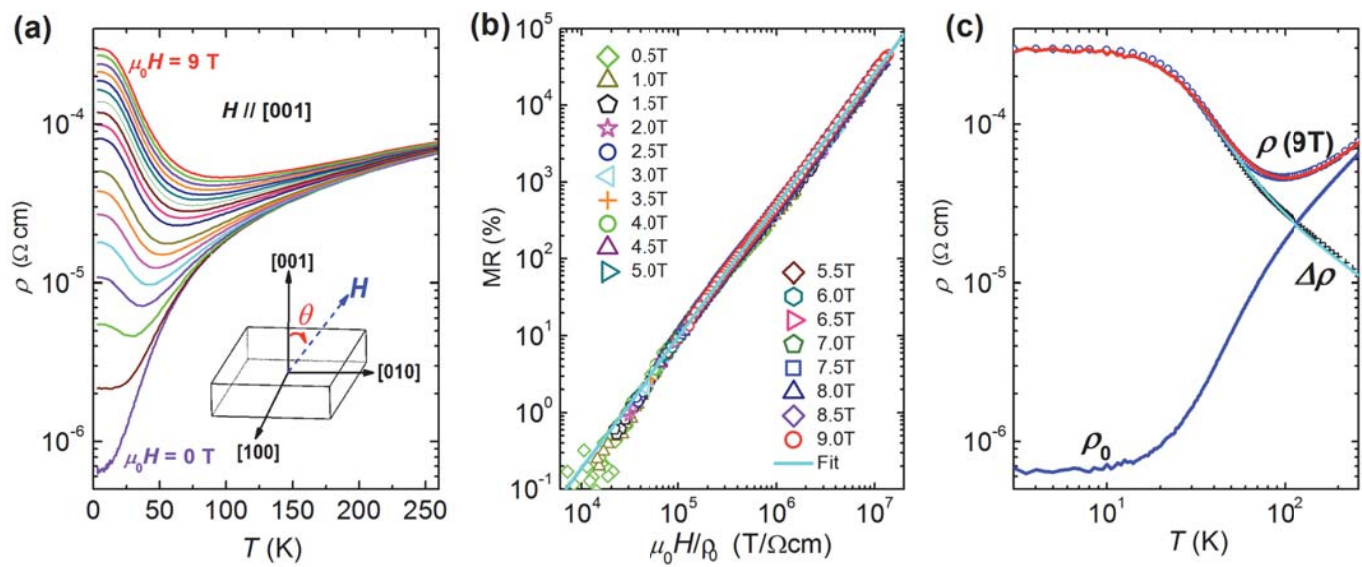

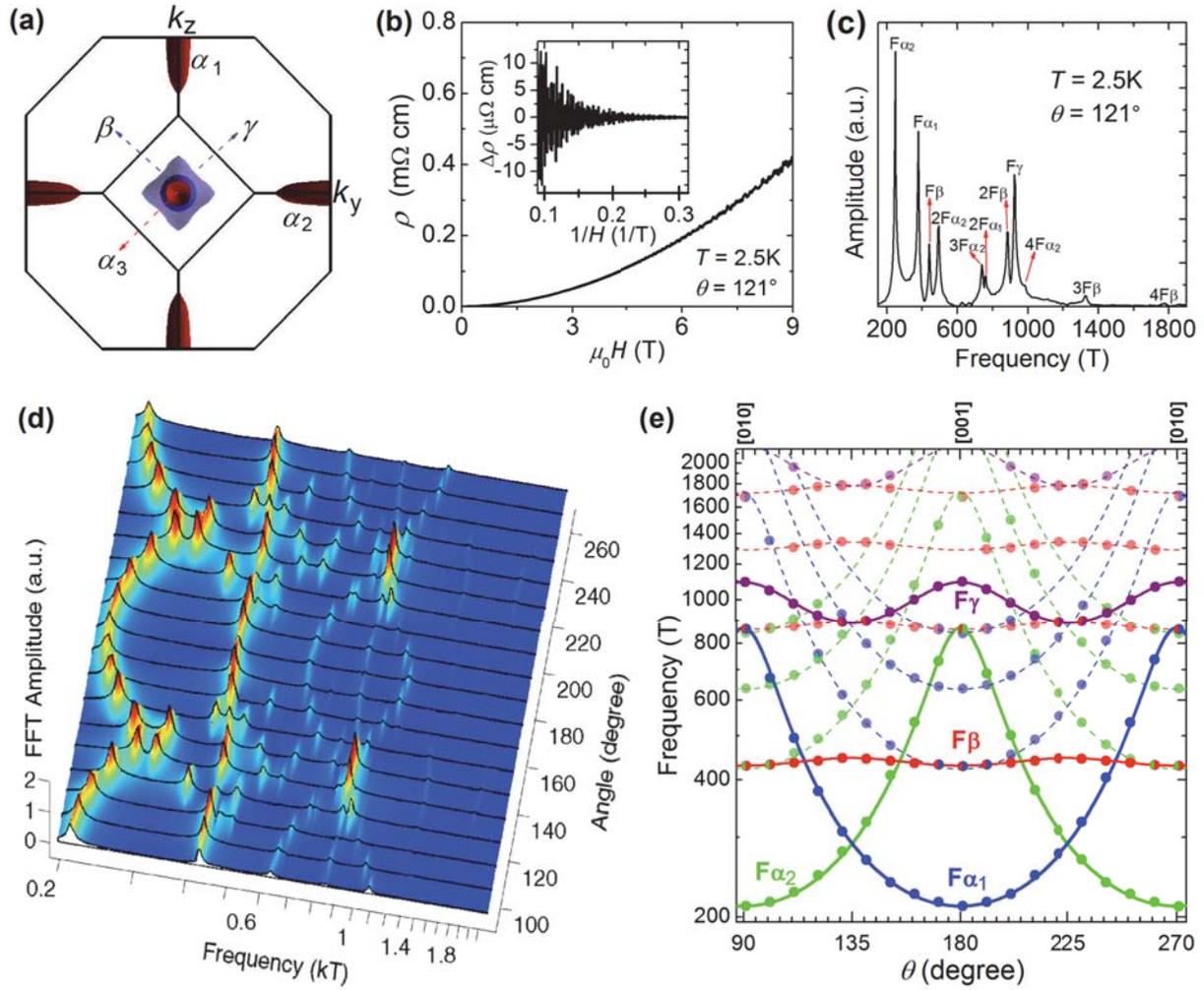

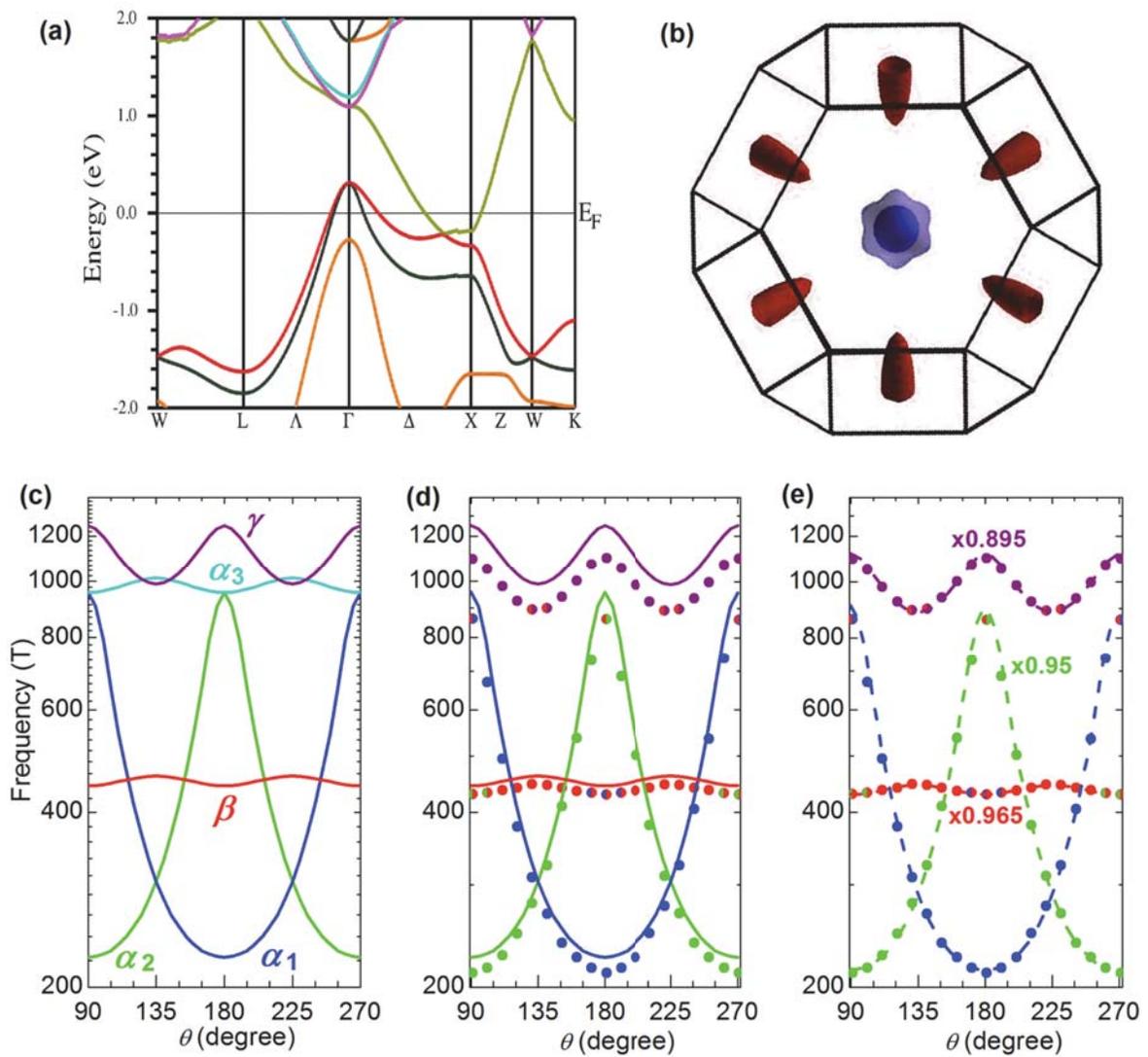

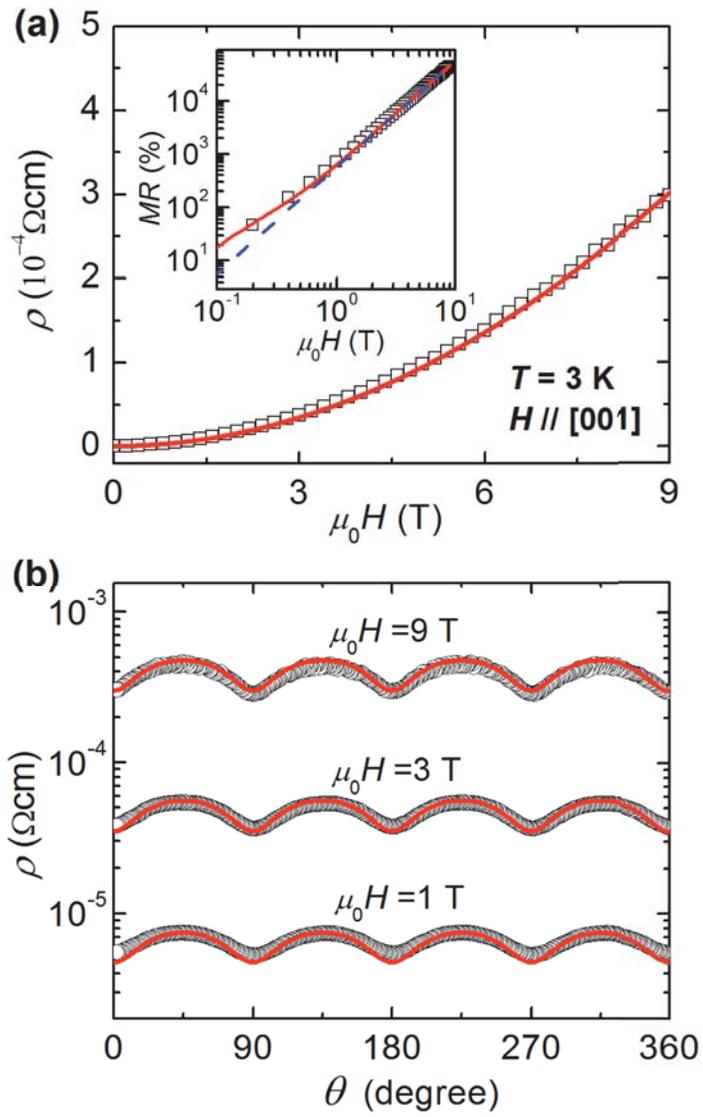

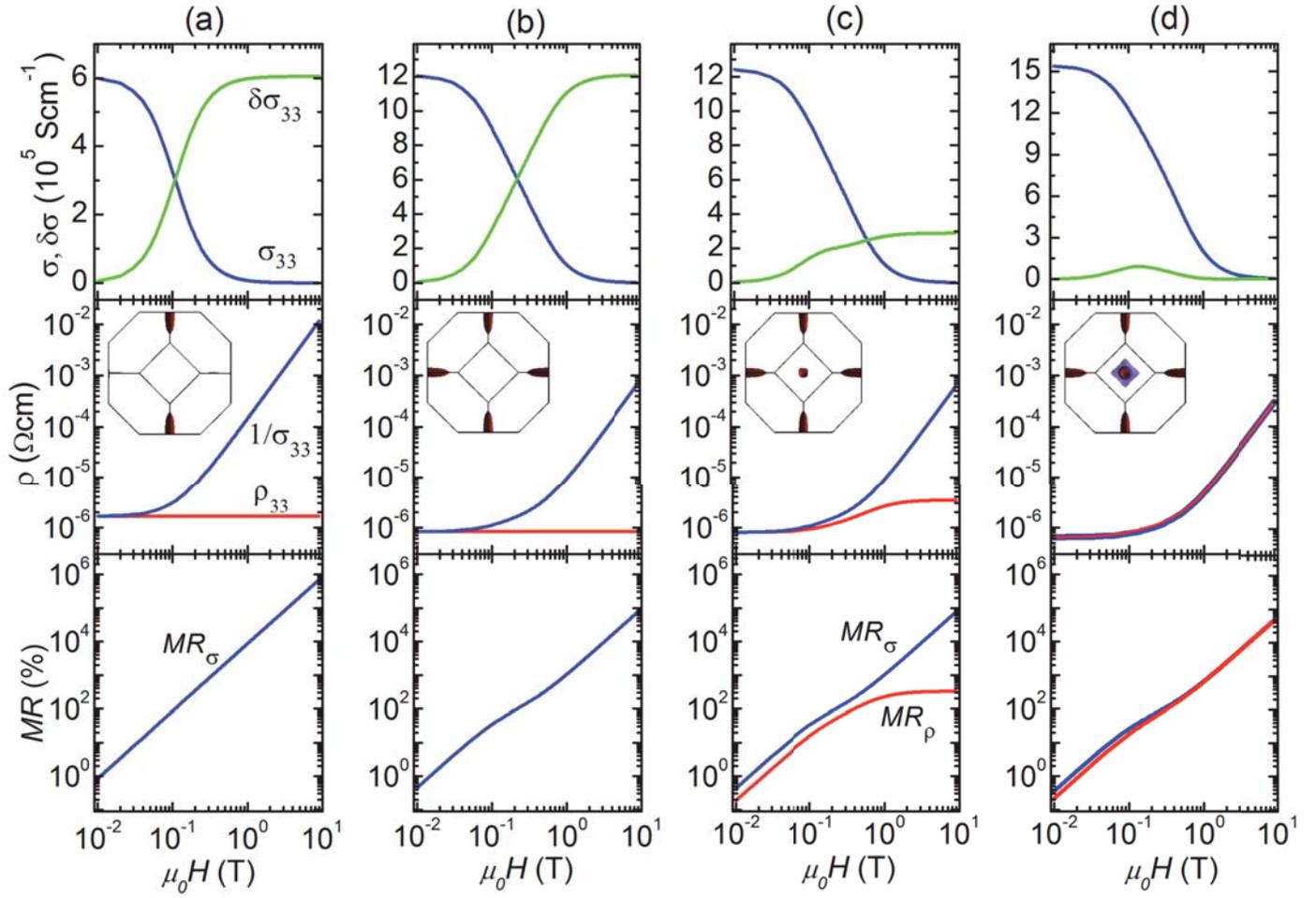

**FIGURE 6**

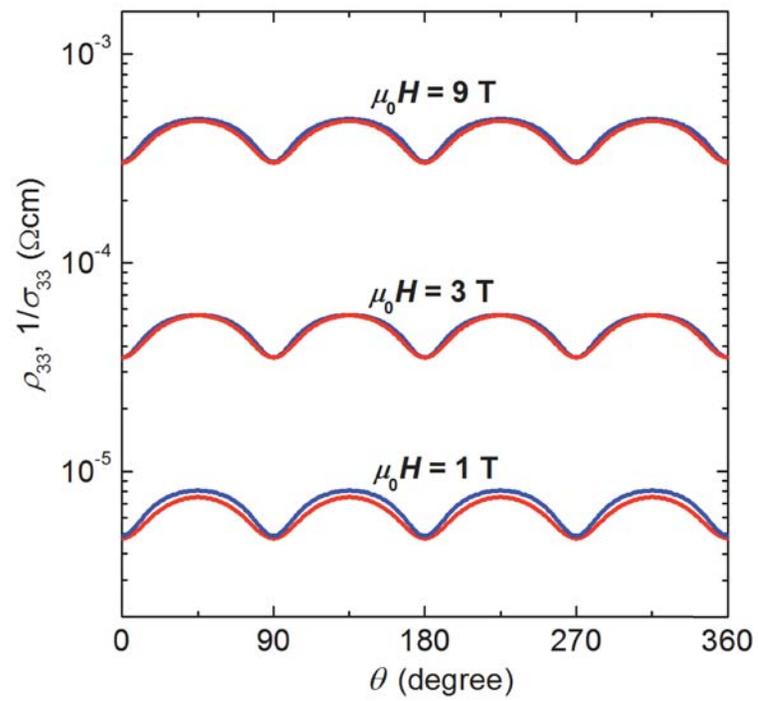

**FIGURE 7**

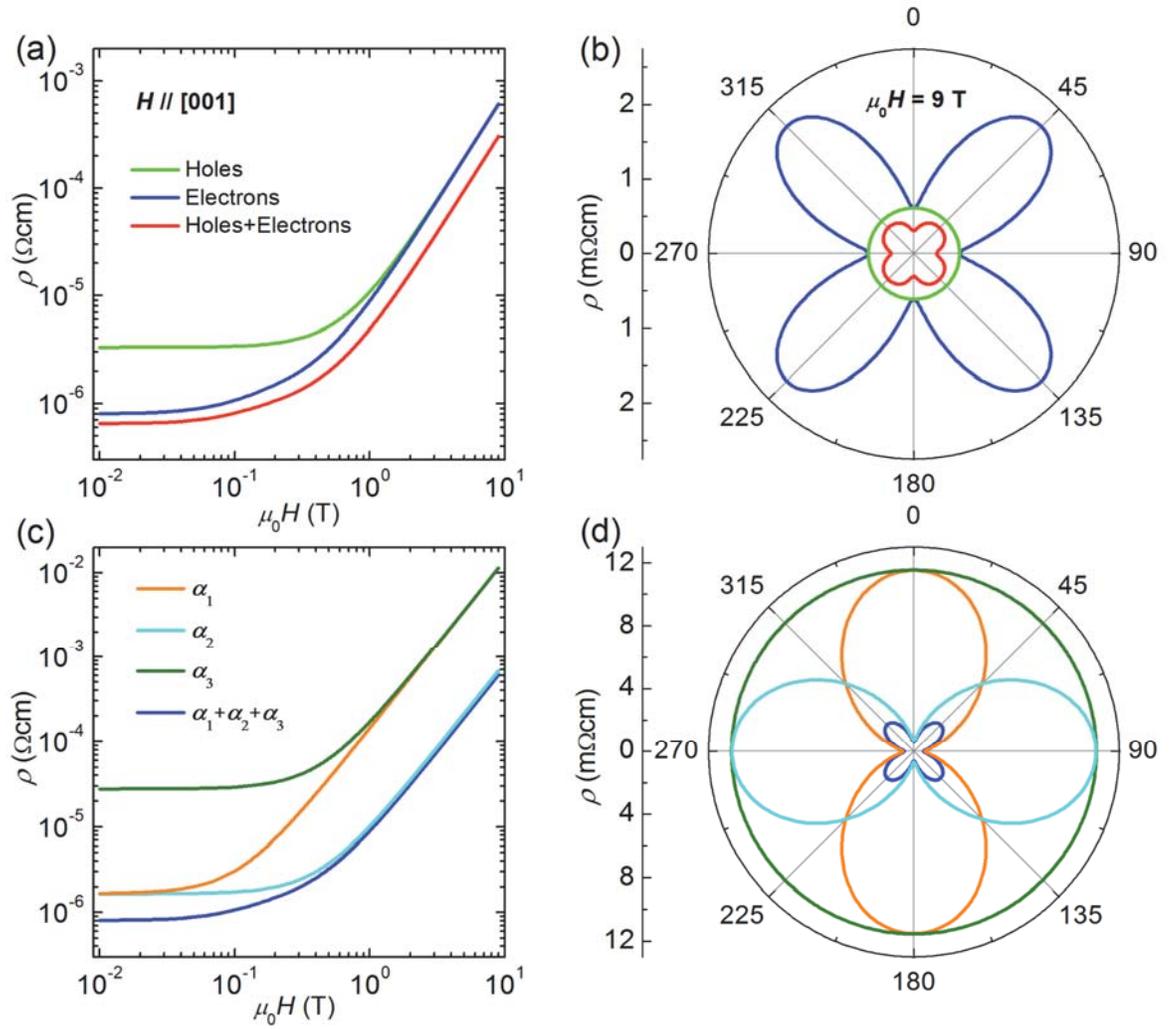

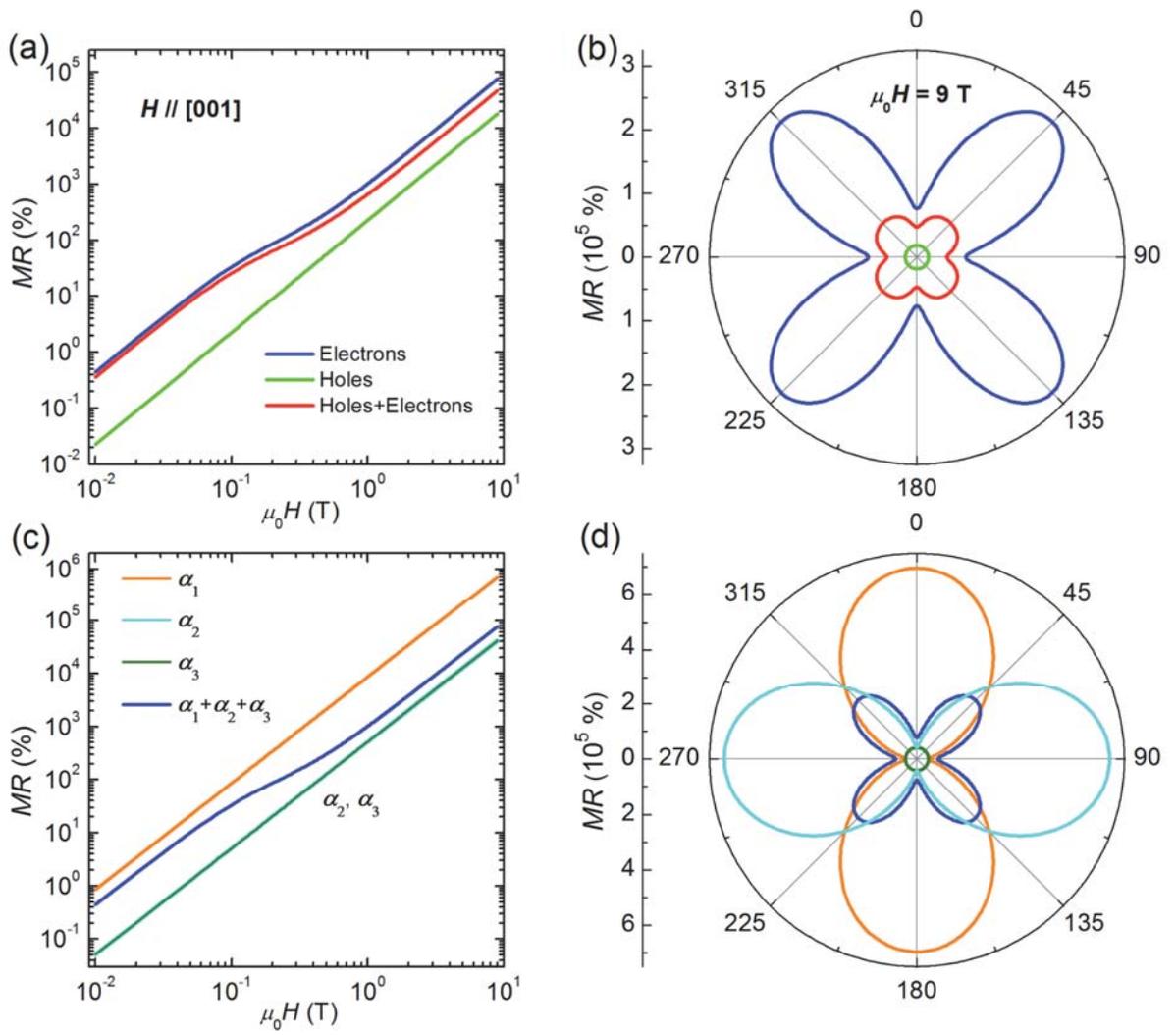